\documentstyle[12pt,preprint]{aastex}
\newcommand{\soe}{1RXS~1708$-$4009}
\newcommand{\tfn}{1E~2259+586}

\newcommand{\rxte}{{\it RXTE}}
\def\nuddd{\ifmmode\stackrel{\bf \,...}{\textstyle \nu}\else$\stackrel{\,...}{\textstyle \nu}$\fi}
\def\nudddd{\ifmmode\stackrel{\bf \,....}{\textstyle \nu}\else$\stackrel{\,....}{\textstyle \nu}$\fi}

\begin{document}

\title{A Second Glitch from the ``Anomalous'' X-ray Pulsar 1RXS J170849.0$-$4000910}

\author{
V. M. Kaspi\altaffilmark{1,2,3} and F. P. Gavriil\altaffilmark{1,4}
}

\altaffiltext{1}{Department of Physics, Rutherford Physics Building,
McGill University, 3600 University Street, Montreal, Quebec,
H3A 2T8, Canada}
\altaffiltext{2}{Department of Physics and Center for Space Research,
Massachusetts Institute of Technology, Cambridge, MA 02139}
\altaffiltext{3}{Canada Research Chair, Steacie Fellow, CIAR Fellow; vkaspi@physics.mcgill.ca}
\altaffiltext{4}{gavriil@physics.mcgill.ca}

\begin{abstract}
We report on 5.4~yr of phase-coherent timing, using the {\it Rossi
X-ray Timing Explorer,} of the X-ray pulsar 1RXS J170849.0$-$4000910
(\soe), a member of the class
known as ``anomalous X-ray pulsars.'' This object exhibited a
rotational glitch in 1999.
Here we report a second much larger rotational glitch which occured
$\sim$1.5~yr after the first.  We show that the recoveries from the two
glitches are significantly different, with the first showing only a
possible slow, approximately linear recovery, while the second showed a
nearly complete recovery on a time scale of $\sim$50~days.  The
approximately exponential recovery time scale of the second glitch is
similar to that seen recently in 1E~2259+586 at the time of a major
outburst.  This suggests \soe\ undergoes similar bursting behavior,
although with our sparse observations we have detected no other
evidence for bursts from this source.
\end{abstract}


\section{Introduction}
\label{sec:intro}

The exotic class of pulsars collectively known as ``anomalous X-ray
pulsars'' \citep[AXPs][]{ms95,vtv95} has recently been shown
\citep{gkw02,kgw+03} to exhibit short, hard-spectrum X-ray bursts that
are very similar to those long seen in another equally exotic class
of objects, the soft gamma repeaters (SGRs).  The properties of the
latter class have been well described by the magnetar model, in which
the objects are isolated, young neutron stars that are powered by the
decay of an enormous stellar magnetic field \citep{td95,td96a,kds+98}.  
In exhibiting bursts, AXPs have unified the two
classes of object, and provided strong support that the magnetar model
applies to both.  However, what, if anything, physically distinguishes the two
classes is still unknown.  Observationally, AXPs appear to
be the less active of the two source classes, with only one major bursting
episode ever detected (and that, fortutously), compared with many
seen from SGRs \citep[see][for a review]{hur00}.

1RXS J170849.0$-$4000910 (which, for brevity, we henceforth refer to
as \soe) is an 11-s X-ray pulsar discovered by \cite{snt+97} in {\it ASCA}
data, and shown to be spinning down like other AXPs by
\cite{ics+99}.  \cite{kcs99} showed that the pulsar is capable of very
stable rotation on the basis of $\sim$1.4~yr of {\it Rossi X-ray Timing
Explorer} ({\it RXTE}) data that permitted phase-coherent timing using
a simple spin-down model with only two free parameters.  \cite{klc00}
reported the discovery of the first rotation glitch seen in an AXP in
data from \soe, and showed that the glitch had properties (e.g. fractional
frequency change $\Delta \nu /
\nu \simeq 6 \times 10^{-7}$) similar to those seen in the Vela
radio pulsar and other radio pulsars like it
\citep[e.g.][]{sl96,wmp+00}.  The rotational stability away from
the glitch and the similarity of the glitch properties with those seen
in radio pulsars lent supporting evidence for the magnetar model.
However, no activity beyond that single glitch has been reported
for the pulsar.

Interestingly, the major bursting episode observed with \rxte\ from a different
AXP, \tfn, was accompanied by a significant rotational glitch, having $\Delta \nu /
\nu = 4 \times 10^{-6}$, of which a fraction $\sim$0.25 recovered on
a time scale of $\sim$50~day \citep{kgw+03,wgk+03}.  Also observed were
short-lived pulse profile variations and an X-ray flux
enhancement.  This event occurred in spite of the pulsar having shown
stable rotational behavior in over 5~yr of previous
\rxte\ monitoring \citep{kcs99,gk02}, although the long-term past
record hinted at episodes of activity \citep{ikh92,cso+95,bs96,hh99}.
This behavior suggests that glitching may be a typical characteristic
of AXPs during outbursts, and/or vice versa.  However, with only one
such event seen, such conclusions are tentative at best.

Here we report on continued \rxte\ monitoring of \soe.  Our data set
now extends over 5.4~yr.  In this interval, in addition to the glitch
reported by \citet{klc00}, we have detected a second glitch
$\sim$1.5~yr later.  Accounting for both glitches in a phase-coherent
timing solution that extends over the full data set, 
we find phase residuals of $<$2\% of the pulse period.  We compare the
properties of the two glitches seen in \soe, and show that their
recoveries are quite different, with the second more closely resembling
that seen in \tfn\ in outburst.  This suggests that \soe\ may undergo
similar bursting activity which has gone unseen by our sparse
sampling.  

Note that a subset of the data reported here has recently been
analyzed as well by \citet{dis+03}, who reach similar though 
slightly different conclusions.  We briefly discuss the differences 
between their results and ours.

\section{Observations and Analysis}
\label{sec:obs}

The {\it RXTE} observations described here are a continuation of those
reported by \citet{kcs99,klc00} and \citet{gk02}.  We refer the reader to those
papers for details of the analysis procedure.  Briefly, all
observations were obtained with the Proportional Counter Array
\citep[PCA][]{jsss+96}, with events selected in the energy range 2.5--5.4~keV to
maximize the signal-to-noise ratio.  The data we report on were obtained
roughly monthly from 1998 January through 2003 May and were analyzed
using software designed to handle raw spacecraft telemetry packet
data.  Data were binned at 31.25-ms time resolution and reduced to the solar
system barycenter using the JPL DE200
ephemeris.  Time series were folded at the nominal pulse period using
64 phase bins.  Folded profiles were cross-correlated in the
Fourier domain with a high signal-to-noise-ratio average profile in
order to determine an average pulse arrival time.  Resulting arrival
times were analyzed using the {\tt TEMPO} pulsar timing software
package\footnote{http://pulsar.princeton.edu/tempo} 
in order to derive a fully phase-coherent timing ephemeris over the entire
data set.  Phase coherence over the glitches is accomplished under the
assumption of zero phase jump at the time of the glitch -- any non-zero
phase jump would suggest an unphysically large torque on the star. 

We assume a glitch model of the form observed for radio pulsars \citep[e.g.][]{sl96},
\begin{equation}
\nu(t) = \nu_0(t) + \Delta \nu + \Delta \dot{\nu} t + \Delta \nu_d {\rm e}^{-t/t_d},
\label{eq:glitch}
\end{equation}
where $\nu_0(t)$ is the pre-glitch spin ephemeris, $\Delta \nu$ and $\Delta \dot{\nu}$
are the permanent changes in $\nu$ and $\dot{\nu}$, and $\Delta \nu_d$ is the change
in $\nu$ that recovers exponentially on a time scale $t_d$.
Best-fit model parameters are given in Table~\ref{ta:parms}.
Figure~1 (top panel) shows the model long-term frequency
history of the pulsar.  The spin-ups, though small compared to
the overall spin-down behavior, are just visible to the eye.
Figure~1 (bottom panel) shows the same model with the linear
trend removed, and the observed frequencies superimposed.
These frequencies were determined using the same arrival
time data, but by doing local fits for frequency using
3--4 arrival times per plotted point.

The first glitch (glitch 1) is modelled as a simple step
in $\nu$ and $\dot{\nu}$ \citep{klc00}, i.e. has $\Delta \nu_d = 0$ in
Equation~\ref{eq:glitch}.
The second glitch (glitch 2), however, as can be seen from Table~\ref{ta:parms},
has $\Delta \nu_d >> \Delta \nu$.  A fraction
$Q \equiv \Delta \nu_d / (\Delta \nu + \Delta \nu_d) = 0.96 \pm 0.11$ of
the total frequency jump recovered following the glitch.  This is
consistent with a full recovery, and contrasts with glitch 1, for which $Q=0$.

Furthermore, no significant change in $\dot{\nu}$ occured at glitch 2,
unlike at glitch 1.  We set a 3$\sigma$ upper limit on a change in
$\dot{\nu}$ relative to the long-term average of $< 2 \times
10^{-16}$~Hz~s$^{-1}$.  \citet{dis+03} report a
significant change in $\dot{\nu}$ at glitch 2, at first glance at odds
with what we find.  This is not because of their smaller data set
post-glitch 2.  We can reproduce their result when we fit for
$\ddot{\nu}$ in the inter-glitch interval.  \citet{dis+03} are
reporting an {\it instantaneous} change in $\dot{\nu}$ at the glitch
epoch, while we report the change of the long-term averages.  We
believe the latter is more physically relevant, since the former can be
contaminated (as is the case here) by noise.  Indeed, that we find no
change in $\dot{\nu}$ post-glitch 2 would have to be a coincidence
otherwise.  Reporting the instantaneous change in $\nu$ is appropriate
however, because the evolution of $\nu$ is dominated by the
deterministic spin-down.

Phase residuals following removal of the model given in the Table are
shown in Figure~2.  The RMS residual is 0.019$P$, comparable to that
seen for young radio pulsars \citep[e.g.][]{antt94}.  However, there
are clearly unmodelled features in the residuals.  The data pre-glitch
1 are well described by a simple model including only $\nu$ and
$\dot{\nu}$.  There is marginal evidence for $\ddot{\nu}$, but as
discussed by \citet{kcs99}, its statistical significance is only
$4\sigma$, with removal of the very first data point giving
$2.6\sigma$.  If real, this term could be due to timing noise or to
recovery from a glitch that occurred prior to the commencement of the
observations.  The residuals between glitches, however, definitely show
a significant $\ddot{\nu}$ as reported by \citet{gk02}.  In that
analysis, it was suggested that this could be recovery from a glitch.
However, it could also be a manifestation of timing noise since the
magnitude of the observed $\ddot{\nu}$ is comparable to that detected
pre-glitch 1.  The occurrence of glitch 2 precludes answering the
question of the origin of the inter-glitch $\ddot{\nu}$.
The residuals post-glitch 2, though overall described well by
Equation~\ref{eq:glitch}, show small but significant deviations
from that model.  Initially, the recovery appears to be slightly slower than
the exponential model, resulting in the relatively large residual
just post-glitch.  The longer-term post-glitch-2 residuals cannot be
modelled well by any polynomial with fewer than 8 parameters.
We can, however, model them reasonably well with only 4 parameters, if we
ignore the glitch and consider only post-glitch data.  That
ephemeris is provided in Table~\ref{ta:altparms}, although we believe
it is unlikely to have precise long-term predictive power.
We note that some young radio pulsars do not show simple exponential
recoveries, but rather multiple recovery time scales \citep[e.g.][]{lsp89}.

Because of the bursting behavior seen in two other AXPs, we searched
all PCA data obtained for \soe\ for bursts.  This was done by creating
binned time series for each PCA proportional counter unit separately,
using time resolution 31.25~ms.  We used events in the energy range
2--20~keV.  We used this wider energy range for the burst search compared to
that used in the timing because of the observed relatively hard AXP
burst spectra compared to the pulsed emission spectrum
\citep{gkw02,kgw+03,gkw+03}.  A total of 310.5~ks of PCA exposure was
searched, for the same time span as for the timing analysis.  The burst
search procedure used is described in detail by \citet{gkw+03}.
Briefly, large excursions from a local mean count rate existing in all
PCUs were flagged using a Poissonian probability discriminator.  No
statistically significant bursts were found in any data for \soe.  The
upper limits on burst fluxes and fluences depend strongly on the
(varying) local background rate, the burst morphology and the burst
spectrum.  Our sensitivity to bursts for \soe\ is
a factor of $\sim$3 below that for \tfn\ (as estimated from the
average total PCA count rates from both sources in the 2--20~keV band),
and the faintest bursts detected for \tfn\ have fluence $\sim 3\times
10^{-11}$~erg~cm$^2$ in the 2--10~keV band \citep{gkw+03}.

We also searched for pulsed flux and pulse morphology changes over the
5.4-yr time span, and especially just after the glitch epochs.  Both
were done with data in the 2--10~keV band and using the methods
detailed by \citet{gk02}.  The pulsed flux time series over the 5.4-yr
span is similar in nature to that shown by \citet{gk02}, i.e. stable at
the 20--30\% level.  Specifically, there is no evidence for enhanced
pulsed flux post-glitch 2.  Similarly, we find no evidence for large
pulse profile changes at any epoch in our data set, including
immediately post-glitch.  However, our standard analysis procedure is
not optimized for detection of low-level changes.  We will present a
more detailed pulse profile analysis elsewhere.  We note that
\citet{dis+03} claim evidence for low-level pulse profile variations by
comparing average profiles of data pre-glitch 1, between glitches, and
post-glitch 2.  However, those authors did not consider whether such
low-level changes, occur in intervals that are not separated by glitches.  
Hence if real, the changes they found may be unrelated to the spin-up events.

\section{Discussion}
\label{sec:discussion}

The two glitches observed in \soe\ exhibit very different recoveries,
as discussed in \S\ref{sec:obs} above.  Specifically, glitch 2 was
dominated by a frequency jump that recovered exponentially on a
$\sim$50~day time scale, while glitch 1 showed no such decay.
Interestingly, the recovery seen in glitch 2 is very similar to that
observed in \tfn\ during its 2002 June outburst
\citep{kgw+03,wgk+03}.  For \tfn, however, $Q \simeq 0.25$
\citep{wgk+03}, different from the $Q \simeq 1$ seen for \soe.
We note that although the first glitch was much smaller than the second,
its long-term effect on the rotation of the pulsar is probably much greater, 
mainly because of its apparently permanent change in $\dot{\nu}$.

The similarity of glitch 2 for \soe\ with that of \tfn\ during outburst
suggests that the former may have been accompanied by bursting behavior
as well.  Such bursts would have gone undetected because of the absence
of any PCA observations near the glitch epoch.  Indeed, the detection of
the bursting from \tfn\ was fortuitous, having lasted only a few
hours.  However, in the latter, the pulsed flux increased by over an
order of magnitude at the time of the outburst and decayed on a time scale of
weeks, remaining over a factor of two brighter for $\sim$3~weeks, but not
returning back to its pre-burst value even after a year
\citep{wgk+03}.  The time between our best estimate of the glitch~2
epoch (Table~\ref{ta:parms}) and the next observation of \soe\ was
3~weeks, possibly long enough for a pulsed flux
enhancement or major pulse profile change to have decayed beyond
detectability.  Hence the observations cannot rule out an unseen outburst
at the time of the \soe\ glitches.

It is interesting to compare the glitching properties of
\soe\ with those of the better studied radio pulsars.  Glitches occur
predominantly in radio pulsars having characteristic ages $\sim 10^3 -
10^5$~yr.  Hence the glitches of \soe\ support its youth, independent of
its characteristic age.  \citet{ml90} introduced a glitch ``activity
parameter'' for radio pulsars, defined as the sum of all frequency
increments in a data span, divided by that data span duration.  This
quantity, admittedly on the basis of only two glitches in \soe,
$\sim 3\times 10^{-15}$~s$^{-2}$, is in the approximate middle of the
distribution of glitch activities for radio pulsars \citep{wmp+00}, and
two orders of magnitude smaller than that for observed for the Vela
pulsar.  The frequency of glitches in \soe, again, tentatively
estimated from only two glitches, is $\sim$0.4~yr$^{-1}$, again in the
approximate middle of the distribution for young radio pulsars, and
substantially below the record holder, PSR~J1341$-$6220
\citep{kmj+92,wmp+00}.  The difference in recovery behavior between the
two glitches seen in \soe\ is not without precedent in radio pulsars;
for example, both PSRs J1731$-$4744 and J1803$-$2137 have shown order
of magnitude differences in $Q$-values for different glitches
\citep{wmp+00}.  However, two \soe\ glitch properties are unusual among
radio pulsar glitches.  The 50-day recovery time scale of glitch 2
is relatively short, although shorter recoveries have been seen for the 
Crab and Vela pulsars.  In addition, the large $Q$ of glitch~2 is unprecedented
in radio pulsar glitches, the bulk of which are in the range 0--0.5.

In the context of vortex unpinning models \citep{ai75}, glitches arise
when vortex lines in the neutron-star crustal superfluid unpin
from crustal lattice nuclei, then repin and transfer angular momentum
from the faster rotating superfluid to the crust and coupled stellar
core.  \citet{acp89} identified two regimes of glitch recovery: a
linear regime in which a small superfluid-crust angular velocity lag
suffices to unpin vortices because of weak pinning and for which the
glitch decays exponentially, and a non-linear regime in which
large lags are necessary and the glitch decays slowly or not at all.
\citet{acp89} suggested that hotter, hence younger neutron stars would
exhibit predominantly linear-regime glitches, while colder, hence older
neutron stars non-linear glitches.  There is some evidence for this from
studies of radio pulsar glitches \citep{wmp+00,lsg00}.  The
detection of both types in \soe\ indicates, in the context of this
model, that both strong and weak vortex line pinnings are present in
the crust.  This is true of many radio pulsars.  For a lengthier
discussion of glitch models, see \citet{dis+03}.

If glitch 2 in \soe\ was accompanied by unobserved bursts and
persistent emission changes as observed for \tfn, the event could have
resulted from a sudden event in the stellar crust, such as a crustal
fracture, which simultaneously affected both the superfluid interior
and the magnetosphere.  Such a crustal fracture is an expected result of stresses
due to the decay of the large stellar magnetic field.   In this scenario,
the fracture both triggered the vortex-line unpinning, and simultaneously
shifted magnetic footpoints, resulting in a magnetospheric reconfiguration.
Searches in future data from \soe\ for behaviors like those seen for 
\tfn\ during its 2002 June outburst are clearly warranted.

We thank R. Manchester for discussions regarding glitch-fitting
with {\tt TEMPO}, J. Swank and the \rxte\ scheduling team, and
the referee A. Lyne for helpful comments.
This work was supported in part by an NSERC Discovery Grant and
Steacie Supplement, FCAR/NATEQ grants, a Canadian Institute for
Advanced Research Fellowship, the Canada Research Chair Program, 
and NASA/LTSA.


\clearpage
\begin{deluxetable}{l|c}
\tablecaption{Spin Parameters for \soe. \label{ta:parms}}
\tablewidth{320pt}
\tablehead{\colhead{Parameter} & \colhead{Value}}
\startdata
Spin Frequency, $\nu$ (Hz) & 0.0909136408(13) \\
Spin Frequency Derivative, $\dot{\nu}$ (Hz s$^{-1}$) & $-1.5681(4) \times 10^{-13}$  \\
Spin Period, $P$ (s) & 10.99944949(15) \\
Spin Period Derivative, $\dot{P}$ & $18972(5) \times 10^{-15}$ \\
Epoch (MJD) & 51459.0000 \\\hline
Glitch 1 & \\\hline
$\Delta \nu$ (Hz) & $4.99(18) \times 10^{-8}$ \\
$\Delta \dot{\nu}$ (Hz s$^{-1}$) & $-0.0157(7) \times 10^{-13}$ \\
Glitch 1 Epoch (MJD) & 51444.601 \\\hline
Glitch 2 & \\\hline
$\Delta \nu$ (Hz) & $1.28(25) \times 10^{-8}$ \\
$\Delta \dot{\nu}$ (Hz s$^{-1}$) & $-0.0001(7)\times 10^{-13}$ \\
$\Delta \nu_d$ (Hz) & $37(3) \times 10^{-8}$ \\
$t_d$ (days) & 50(4) \\
Glitch 2 Epoch (MJD) & 52014.177 \\\hline
RMS Timing Residual (ms) & 213 \\
Number of Arrival Times & 81 \\
Start Observing Epoch (MJD) & 50826 \\
End Observing Epoch (MJD) & 52786 \\
\enddata
\tablecomments{Numbers in parentheses represent 1$\sigma$ uncertainties
in the last digit quoted.}
\end{deluxetable}

\clearpage
\begin{deluxetable}{l|c}
\tablecaption{Alternative Spin Ephemeris for \soe\ Post Glitch 2. \label{ta:altparms}}
\tablewidth{0pt}
\tablehead{\colhead{Parameter} & \colhead{Value}}
\startdata
$\nu$ (Hz) & 0.0908958219(28) \\
$\dot{\nu}$ (Hz s$^{-1}$) & $-1.540(7) \times 10^{-13}$ \\
$\ddot{\nu}$  (Hz s$^{-2}$) & $8.9(1.0) \times 10^{-22}$ \\
\nuddd\  (Hz s$^{-3}$) & $8.6(8) \times 10^{-29}$ \\
\nudddd\ (Hz s$^{-4}$) & $3.46(26) \times 10^{-36}$ \\
Epoch (MJD) & 52766.000 \\
RMS Timing Residual (ms) & 184 \\
Number of Arrival Times & 39 \\
Start Observing Epoch (MJD) & 52035 \\
End Observing Epoch (MJD) & 52786 \\
\enddata
\tablecomments{Numbers in parentheses represent 1$\sigma$ uncertainties
in the last digit quoted.}
\end{deluxetable}

\clearpage
\begin{figure}
\plotone{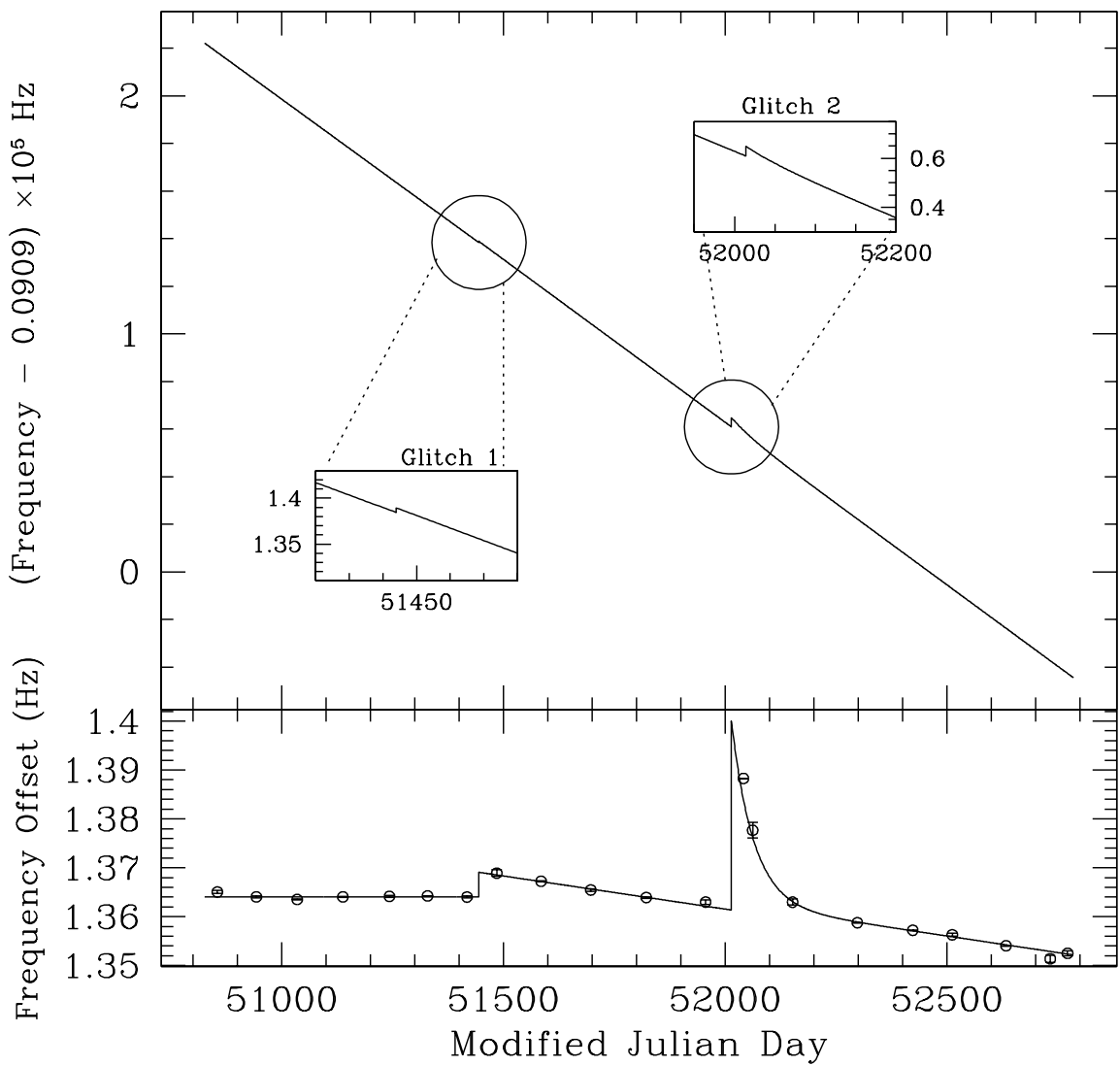}
\newpage
\figcaption{(Top) Model frequency history for 5.4~yr of \protect\rxte\ monitoring
(see Table 1).
The glitches, superimposed on the overall spin down, are indicated
by circles.  The insets show the glitch data on an expanded scale.  The
recovery of glitch 2 is discernible by eye.
(Bottom)  The same model but with the linear trend removed, and frequencies referred to epoch
MJD 51459.000.  The two glitches are obvious.  The data points, whose uncertainties
are shown but which are generally smaller than the size of the point, 
were determined by doing local frequency fits using 3--4 arrival times per epoch.}
\label{fig:freq}
\end{figure}

\clearpage
\begin{figure}
\plotone{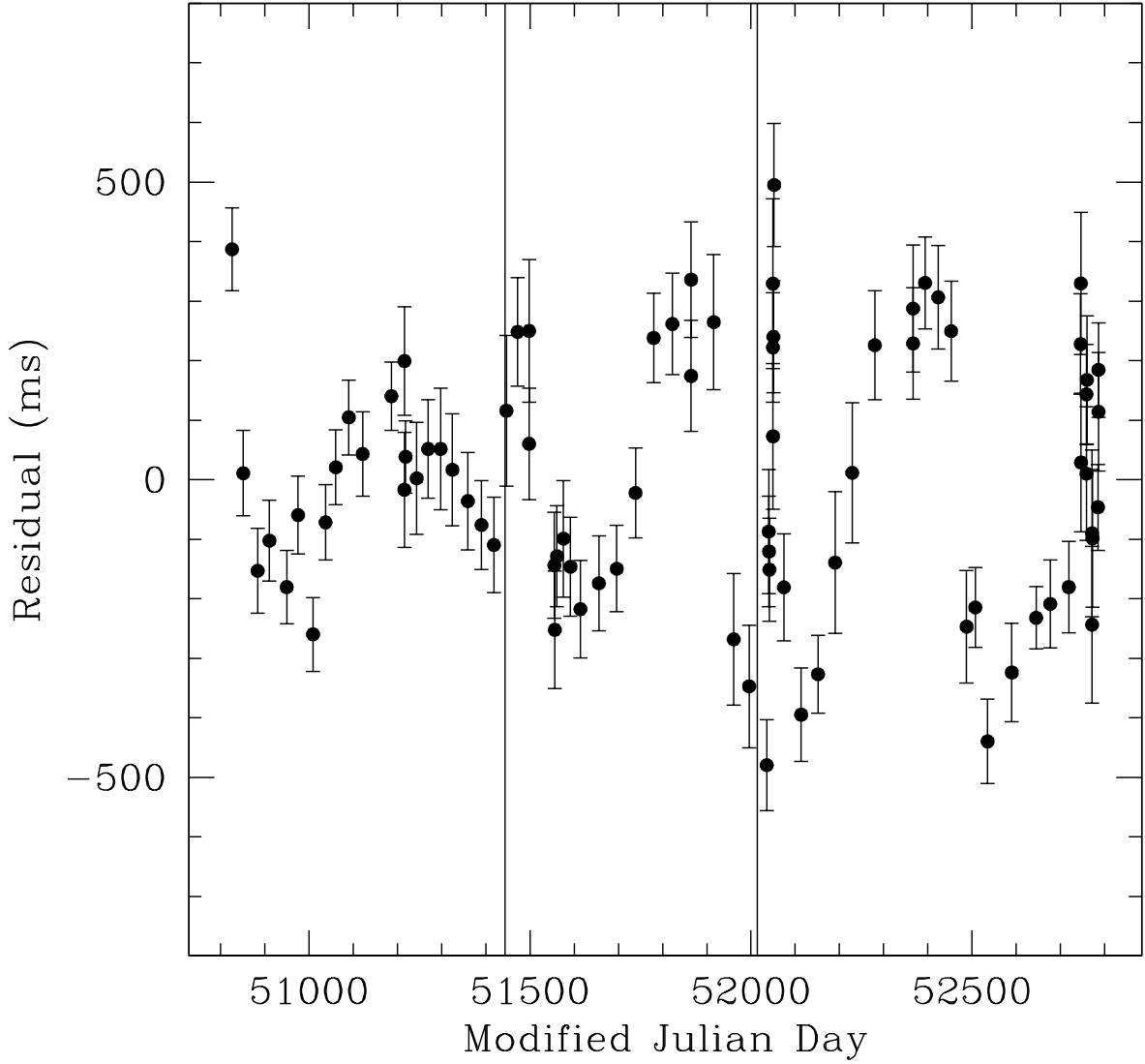}
\newpage
\figcaption{Phase residuals for 5.4~yr of phase-coherent timing of
\protect\soe, with
the model shown in Table~\protect\ref{ta:parms} removed.  The best-fit glitch
epochs are indicated with vertical lines.  The RMS
residual is 0.019 periods.  Residuals pre-glitch 1 show marginal evidence for
$\ddot{\nu}$ as discussed by \protect\citet{kcs99}.  Residuals between
glitches show a significant $\ddot{\nu}$ which is consistent with standard
recovery of the change in $\dot{\nu}$ as discussed in \protect{\citet{gk02}},
but which may also be a manifestation of timing noise.  Residuals post-glitch 2
show significant structure that indicates the short-time scale glitch response 
was slightly slower than exponential.  The post-glitch 2 residuals can also
be reasonably described by a polynomial fit with 4 parameters 
(see text and Table~\protect{\ref{ta:altparms}}).
}
\label{fig:resid}
\end{figure}

\end{document}